\documentclass[twocolumn, superscriptaddress,nofootinbib, floatfix, amsfonts, aps]{revtex4-1}
\usepackage{amsmath}
\usepackage{bm}
\usepackage{graphicx}
\usepackage{mathrsfs}
\usepackage{footmisc}
\usepackage{multirow}
\usepackage{graphicx}
\usepackage{booktabs, ctable}
\usepackage{soul,xcolor}
\usepackage{xcolor, ulem, color, framed}
\usepackage{amssymb}

\usepackage[bookmarks,bookmarksnumbered]{hyperref}
\hypersetup{colorlinks = true,
            linkcolor = blue,
            anchorcolor =red,
            citecolor = blue,
            pdfauthor=author}

\begin{document}

\setstcolor{red}

\title{Probe the color screening in proton-nucleus collisions with complex potentials}
\author{Liuyuan Wen}
\affiliation{Department of Physics, Tianjin University, Tianjin 300350, China}

\author{Xiaojian Du}
\affiliation{Fakult\"at fur Physik, Universit\"at Bielefeld,
D-33615 Bielefeld, Germany}

\author{Shuzhe Shi}
\affiliation{Center for Nuclear Theory, Department of Physics and Astronomy, Stony Brook University, Stony Brook, New York, 11784, USA.}

\author{Baoyi Chen}
\email{baoyi.chen@tju.edu.cn}
\affiliation{Department of Physics, Tianjin University, Tianjin 300350, China}

\date{\today}

\begin{abstract}
Color screening and parton inelastic scattering modify the heavy-quark antiquark potential in the medium that consists of particles from quantum chromodynamics (QCD), leading to suppression of quarkonium production in 
 relativistic heavy-ion collisions.
Due to small charm/anti-charm ($c\bar{c}$) pair production number in proton-nucleus (pA) collisions, the correlation between different $c\bar{c}$ pairs is negligible, which makes the Schr\"odinger equation viable for tracking the evolution of only one $c\bar{c}$ pair.
We employ the time-dependent Schr\"odinger equation with in-medium $c\bar{c}$ potential  to study the evolution of charmonium wave functions in the hydrodynamic like QCD medium produced in pA collisions.
We explore different parametrizations of real and imaginary parts of $c\bar{c}$ potential
and  calculate the nuclear modification factors ($R_{\rm pA}$) of $J/\psi$ and $\psi(2S)$ in $\sqrt{s_{NN}}=5.02$ TeV energy p-Pb collisions at Large Hadron Collider (LHC). 
Comparing a strong and a weak screening scenario with experimental data in this approach, we arrive at the conclusion that the color screening is weak at temperature close to deconfined phase transition. 
Moreover, the imaginary part of the potential is crucial to describe the experimental data which is consistent with widely studied semi-classical approaches where the dissociation rates are essential.
\end{abstract}
\maketitle
\section{introduction}
Medium of deconfined gluons and light quarks, called the quark-gluon plasma(QGP)~\cite{Bazavov:2011nk}, can be produced in ultrarelativistic heavy-ion collisions (URHICs). 
Charmonium as a bound state of charm ($c$) and anti-charm ($\bar{c}$) quark pair has been proposed to be a clean probe to study the formation of the QGP in heavy-ion collisions~\cite{Matsui:1986dk}. 
Different charmonium states bound by $c\bar{c}$ potential with color screening from in-medium light partons are sequentially melted when the temperature of the QGP medium increases~\cite{Satz:2005hx}.
Besides, charmonium states suffer a direct dissociation from in-medium parton inelastic scattering~\cite{Peskin:1979va,Bhanot:1979vb,Grandchamp:2001pf,Brambilla:2011sg,Brambilla:2013dpa} which corresponds to an additional imaginary part of the $c\bar{c}$ potential~\cite{Laine:2006ns,Brambilla:2008cx}.
In nucleus-nucleus (AA) collisions at the LHC, the initial temperatures of QGP can be far above the dissociation temperature $T_d$ of charmonium ground state $J/\psi$~\cite{Liu:2012zw,Liu:2017qah}.
Most of primordially produced charmonia are dissociated in the medium, and the final production is dominated by the coalescence of abundant charm and anti-charm quarks in the regions where medium local temperature become smaller than the dissociation temperatures of charmonium~\cite{Thews:2000rj,Yan:2006ve,Grandchamp:2002wp}. 
The final spectrum of the charmonia are affected by the charm quark diffusion in medium and their coalescence probability below dissociation temperature~\cite{Zhao:2007hh,Du:2015wha,Zhao:2017yan,Zhao:2018jlw,He:2021zej}. 

In p-Pb collisions at $\sqrt{s_{NN}}=5.02$ TeV, a small deconfined medium is also believed to be generated~\cite{Zhao:2020wcd}, where the medium temperature is slightly above the critical temperature $T_c$ but still below the dissociation temperature $T_d\simeq 2T_c$ of $J/\psi$, which is on the order of its binding energy $T_d\simeq E_{b}$. Consider that the mass and typical momentum of charm quark are larger than charmonium binding energy $m > p > 3T >T_d \simeq E_b$, one can integrate out the hard scale $m$, soft scale $p$ and live with a non-relativistic potential description~\cite{Pineda:1997bj}.

Besides, the recombination of charmonium production becomes negligible in pA collisions~\cite{Chen:2016dke,Du:2018wsj} due to small $c\bar{c}$ production. 
That excludes the contamination from the coalescence of $c\bar{c}$ pair and the correlation between different $c\bar{c}$ pairs which lead to the recombination contribution of charmonium. 
Those features make the Schr\"odinger equation which evolves only one $c\bar{c}$ pair in a potential viable for a quantum description of charmonium in pA collisions. 

With similar considerations, recombination of bottomonia in URHICs is negligible~\cite{Grandchamp:2005yw,Liu:2010ej,Emerick:2011xu,Du:2017qkv,Yao:2020xzw}, and open quantum system (OQS) description of bottomonium evolution in QCD medium which tracks one bottom/anti-bottom ($b\bar{b}$) pair has been carried out by various models in recent years~\footnote{Another consideration for tremendous discussions on bottomonium is due to its heavy mass which makes the relativistic effects negligible (NRQCD)~\cite{Caswell:1985ui,Bodwin:1994jh} and typical momentum relatively larger compared to the binding energy, rendering a potential (pNRQCD)~\cite{Pineda:1997bj}.}. 
The QCD medium environment can be encoded in the Hamiltonian of the heavy quark/antiquark ($Q\bar{Q}$) subsystem as additional terms equivalent to real and imaginary parts of the potential. 
Those studies start with solving Schr\"odinger equation with a complex potential~\cite{Strickland:2011mw,Krouppa:2015yoa,Boyd:2019arx}, to a stochastic potential~\cite{Akamatsu:2011se} and Schr\"odinger-Langevin type equation~\cite{Katz:2015qja}. 
More involved calculations tend to evolve the density matrix of the $Q\bar{Q}$ subsystem with Lindblad formalism~\cite{Brambilla:2016wgg}, whose different terms represent color screening, primordial dissociation and recombination of one pair respectively~\cite{Yao:2018nmy}.
One of such calculations incorporated with quantum trajectory method can be found in~\cite{Brambilla:2020qwo}.

In this paper, we will not go for more complicated quantum treatment as people discussed for bottomonium, but rather solve the Schr\"odinger equation with complex potential.
We parameterized the in-medium temperature dependent complex potential of $c\bar{c}$ dipole according to lattice QCD data. 
Then we evolve a $c\bar{c}$ pair wave-function by solving time-dependent Schr\"odinger in position space with Crank-Nicolson~\cite{crank_nicolson_1947} implicit method.
The final production of $J/\psi$ and $\psi(2S)$ are obtained via projecting the wave-functions of $c\bar c$ dipoles to the charmonium wave-function (from solving time-independent Schr\"odinger equation with vacuum potential) after they leave the hot medium along different trajectories. 
Since the geometric sizes of different charmonium wave-functions are different, $J/\psi$ and $\psi(2S)$ experience different magnitudes of the screening effect and the inelastic collisions with thermal partons. 
This results in different dissociations of $J/\psi$ and $\psi(2S)$. 

Since the suppressions of $J/\psi$ and $\psi(2S)$ are clearly distinguished shown by experimental data, it is essential to employ different scenarios of potential in the Schr\"odinger equation and understand the potential in play behind data, especially their overall suppression and relative suppression.
In order to understand the role the color screening plays, we implement a strong and a weak screening scenario with and without imaginary part of the potential.

This paper is organized as follows. 
In Sec.~\ref{sec.2}, the Schr\"odinger equation model and parametrizations of the heavy quark potential are introduced. Medium evolution which provides the space-time dependent temperature profile is described by hydrodynamic equations. 
In Sec.~\ref{sec.3}, we discuss the application of the model to p-Pb collisions at the LHC energy, $R_{pA}$s of $J/\psi$ and $\psi(2S)$ are calculated with different in-medium potentials and compared with the experimental data. 
We conclude in Sec.~\ref{sec.4}.

\section{Schr\"odinger equation model}
\label{sec.2}
\subsection{Initial distributions}
Heavy quark dipoles are produced in initial parton hard scatterings and then evolve into charmonium eigenstates. 
The momentum distribution of $c\bar c$ dipoles is approximated to be the $J/\psi$ momentum distribution in proton-proton (pp) collisions. 
Therefore in p-Pb collisions, the initial distribution of primordially produced $c\bar c$ dipoles can be obtained through a superposition of the effective pp  collisions~\cite{Chen:2016dke},
\begin{align}
\label{eq-initial}
f_{\Psi}({\bf p},{\bf x}|{\bf b}) &= (2\pi)^3\delta(z)T_\text{p}({\bf x}_T)T_\text{A}({\bf x}_T-{\bf b}) \nonumber \\
&\times 
{\cal R}_{\text g}(x_g,\mu_{\text F},{\bf x}_{\text T}-{\bf b})
{d \bar \sigma^\Psi_{pp}\over d^3{\bf p}},
\end{align}
where $\bf b$ is the impact parameter, 
$\bf x_T$ is the transverse coordinate,
$T_A({\bf x_T})=\int dz\rho_A({\bf x_T},z)$ is the nuclear thickness function, and the nuclear density is taken as Woods-Saxon distribution. 
$T_p({\bf x_T})$ is the proton thickness, where proton density is taken as a Gaussian distribution~\cite{Chen:2016dke}. The 
width of Gaussian function is determined with the proton charge radius $\langle r\rangle_p=0.9$ fm~\cite{Gao:2021sml}. 
The shadowing effect is included with the inhomogeneous modification factor ${\cal R}_\text{g}$~\cite{Vogt:2004dh} for the gluons with the longitudinal momentum $x_g=e^{y}\ E_\text{T}/\sqrt{s_{\text{NN}}}$ and the factorization factor $\mu_\text{F}=E_\text{T}$. The transverse energy and the momentum rapidity are defined as $E_\text{T}=\sqrt{m_\Psi^2+{\bf p}_\text{T}^2}$ 
and $y=1/2\ln((E+p_\text{z})/(E-p_\text{z}))$. The values of the gluon shadowing factor $\mathcal{R}_g$ is obtained with EPS09 model~\cite{Eskola:2009uj}. 
The effective initial momentum distribution ${d \bar \sigma^\Psi_{pp}\over d^3{\bf p}}$ of charmonium in p-Pb collisions have included the Cronin effect~\cite{Cronin:1974zm}. 
Before two gluons fuse into a heavy quark dipole, they obtain additional transverse momentum via multi-scatterings with the surrounding nucleons. 
The extra momentum will be inherited by the produced $c\bar c$ dipole or charmonium states. 
With the random walk approximation, the Cronin effect is included with the modification in the momentum-differential cross section measured in pp collisions, 
\begin{align}
\label{eq-cronin}
{d\bar \sigma_{pp}^{\Psi}\over d^3{\bf p}} = {1\over \pi a_{gN}l}\int 
d^2{\bf q}_T e^{-{\bf q}_T^2\over a_{gN}l} 
{d\sigma_{pp}^{\Psi}\over d^3{\bf p}} 
\end{align}
{
where $l({\bf x}_T)=0.5T_A({\bf x}_T)/\rho_A({\bf x}_T,z=0)$ is the averaged path length of 
the gluon in the nucleus 
travelling through before scattering with the other gluon in the proton to produce 
a heavy quark dipole at the position 
${\bf x}_T$.}
  $a_{gN}$ represents the extra transverse momentum square in a unit of length of nucleons before the fusion process. Its value is taken to be $a_{gN}=0.15\ \rm{GeV^2/fm}$~\cite{Chen:2018kfo}. 
The charmonium distribution in pp collisions have been measured by ALICE Collaboration at 2.76 TeV and 7 TeV~\cite{ALICE:2012vup,ALICE:2011zqe}. 
With these data, we parametrize the normalized $p_T$ distribution of charmonium at $\sqrt{s_{NN}}=5.02$ TeV to be, 
\begin{align}
{dN_{J/\psi}\over 2\pi p_T d{ p_T}} = {(n-1)\over \pi (n-2)
\langle p_T^2\rangle_{pp}}[1+{p_T^2\over (n-2)\langle p_T^2\rangle_{pp}}]^{-n}
\end{align}
where $n=3.2$ and the mean transverse momentum square of charmonium is parametrized as $\langle p_T^2\rangle_{pp}(y)=12.5\times [1- (y/y_{\rm max})^2] \rm{(GeV/c)^2}$ where the maximum rapidity of charmonium is defined with $y_{\rm max}=\ln(\sqrt{s_{NN}}/m_\Psi)$~\cite{Chen:2015iga}. 
$m_\Psi$ is the charmonium mass.

\subsection{Evolution of $c\bar c$ dipoles in the medium}
The heavy quark potential of $c\bar c$ dipole is modified by the hot medium, which affects the evolution of charmonium wave functions~\cite{Kajimoto:2017rel,Guo:2015nsa, Chen:2016vha}. 
Hot medium effects can be included in the Hamiltonian of $c\bar c$ dipoles. 
As a charm quark is heavy compared with the inner movement of charmonium bound states, 
the relativistic effect is ignored when considering the inner structure of a charmonium. 
We employ the time-dependent Schr\"odinger equation to describe the evolution of $c\bar c$ dipole wave functions with in-medium complex potentials. 
Assume the heavy quark-medium interaction is spherical without angular dependence, there is no mixing between charmonium eigenstates with different angular momentum in the wave function of the $c\bar c$ dipole. 
Radial part of the wave function of $c\bar c$ dipole at the center of mass frame is separated as, 
\begin{align}
\label{fun-rad-sch}
i\hbar {\partial \over \partial t}\psi( r, t) = \Big[-{\hbar^2\over 2m_\mu}{\partial ^2\over \partial r^2} +V( r, T) + {L(L+1)\hbar^2\over 2 m_\mu r^2}\Big]\psi(r,t)
\end{align}
where $r$ is the relative distance between charm and anti-charm quarks and $t$ is the proper time in the center of mass frame. $m_\mu=m_1m_2/(m_1+m_2)=m_c/2$ is the reduced mass and 
$m_c$ is charm quark mass. $\psi(r,t)$ is defined to be $\psi(r,t)=r R(r,t)$, where $R(r,t)$ is the radial part of the $c\bar c$ dipole wave function. 
The complete wave function of the $c\bar c$ dipole can be expanded in the eigenstates of the vacuum Cornell potential, $\Psi(r,\theta, \phi)= \sum_{nlm}c_{nlm}R_{nl}(r)Y_{lm}(\theta, \phi)$. $Y_{lm}$ is the spheric harmonics function. $L=(0,1,...)$ is the quantum number of the angular momentum. In the ideal fluid with zero viscosity, heavy quark potential $V(r,T)$ is radial. 
There is no transitions between charmonium eigenstates with different angular momentum $L$. 
The potential depends on local temperature of the medium which is given by hydrodynamic model 
in the next Section.  Radial Schr\"odinger equation 
Eq.~(\ref{fun-rad-sch}) is solved numerically with the Crank--Nicolson method (take natural units $\hbar=c=1$). 
The numerical form of the Schr\"odinger equation is simplified as, 
\begin{align}
  \label{eq-sch-num}
{\bf T}_{j,k}^{n+1}\psi_{k}^{n+1} = \mathcal{V}_{j}^{n}.
\end{align}
Here $j$ and $k$ are the index of rows and columns in the matrix $\bf T$ respectively. 
The non-zero elements in the matrix are, 
\begin{align}
\label{eq-sim-cn}
&{\bf T}^{n+1}_{j,j}= 2+2a+bV_j^{n+1}, \nonumber \\
&{\bf T}^{n+1}_{j,j+1}={\bf T}^{n+1}_{j+1,j}= -a, \nonumber \\
&\mathcal{V}_j^n= a\psi_{j-1}^n +(2-2a-bV_j^n)\psi_j^n +a\psi_{j+1}^n ,
\end{align}
where $i$ is an imaginary number, 
$a= i\Delta t/(2m_\mu (\Delta r)^2)$, and $b=i\Delta t$.
The subscript $j$ and superscript $n$ in $\psi_j^n$ represents 
the coordinate $r_j=r_0 +j\cdot \Delta r$ and the time $t^n=t_0 +n\cdot\Delta t$ 
respectively. $\Delta r$ and $\Delta t$ are the steps of the radius 
and the time in numerical simulation.  
Their values are taken to be $\Delta t=0.001$ fm/c and $\Delta r=0.03$ fm, respectively. 
$t_0$ is the start time of the Schr\"odinger 
equation. 
The matrix ${\bf T}^{n}$ at each time step depends on the in-medium heavy quark potential $V(r,T)$ which will be given later. 

The Schr\"odinger equation Eq.~(\ref{fun-rad-sch}) 
describes the evolution of the wave function of the $c\bar c $ dipole from 
$t\ge t_0$. The initial wave function of the $c\bar c$ dipole is taken to be 
one of charmonium eigenstates. 
After traveling through the hot medium, 
the fractions $|c_{nl}(t)|^2$ 
of each charmonium eigenstate (1S, 1P, 2S, etc) 
in the $c\bar c$ dipoles changes with time. $c_{nl}(t)$ is defined as, 
\begin{align}
c_{nl}(t) &=  \int R_{nl}(r)e^{-iE_{nl} t} \psi(r,t) rdr 
\end{align}
where  
the radial wave function 
$\psi(r,t)$ is given by Eq.~(\ref{eq-sch-num}). 
The ratio of final and initial fractions of a certain charmonium state in one 
$c\bar c$ dipole is written as, 
$R^{\rm direct}(t) ={|c_{nl}(t)|^2\over 
|c_{nl}(t_0)|^2}$. 
In p-Pb collisions, the initial spatial and momentum distributions of 
the primordially produced $c\bar c$ dipoles are given by Eq.~(\ref{eq-initial}). 
After averaging over the positions and  momentum bins of different $c\bar c$ dipoles 
in p-Pb collisions, 
one can get the ensemble-averaged fractions of a certain 
charmonium state in the 
$c\bar c$ dipole $\langle |c_{nl}(t)|^2\rangle_{\rm en}$. 
The 
direct nuclear modification factor 
of charmonium eigenstate ($n,l$) is written as, 
\begin{align}
\label{eq-directRAA}
R_{pA}^{\rm direct}(nl) &={\langle |c_{nl}(t)|^2\rangle_{\rm en}\over 
\langle |c_{nl}(t_0)|^2\rangle_{\rm en}}\nonumber \\
&={\int d{\bf x}_{\Psi}d{\bf p}_{\Psi} |c_{nl}(t, {\bf x}_{\Psi}, 
{\bf p}_{\Psi})|^2{{dN^{\Psi}_{pA}}\over d{\bf x}_{\Psi} d{\bf p}_{\Psi}}
\over 
\int d{\bf x}_{\Psi}d{\bf p}_{\Psi} |c_{nl}(t_0,{\bf x}_0, {\bf p}_{\Psi})|^2
{\overline {dN^{\Psi}_{pA}}\over d{\bf x}_{\Psi} d{\bf p}_{\Psi}}}
\end{align}
where ${\bf x}_{\Psi}$ and ${\bf p}_{\Psi}$ is the position and the total 
momentum of the correlated $c\bar c$ dipole. If without the hot medium effects, 
these correlated $c\bar c$ dipoles are just charmonium eigenstates without 
dissociation. 
${dN_{pA}^{\Psi}\over d{\bf x}_{\Psi}d{\bf p}_{\Psi}}$ is the 
initial spatial and momentum distributions of primordially produced charmonium 
in p-Pb collisions. It is given by Eq.~(\ref{eq-initial}). 
Note that in the denominator, 
${\overline{dN_{pA}^{\Psi}}\over d{\bf x}_{\Psi}d{\bf p}_{\Psi}}$ is 
calculated by Eq.~(\ref{eq-initial}) excluding the cold nuclear matter effects. 
  
After considering the feed-down contributions from excited states, 
one can get the nuclear modification factor of $J/\psi$ (which is 
given in experimental data), 
\begin{align}
\label{eq-promptRAA}
R_{pA}(J/\psi) 
= {\sum_{nl} \langle |c_{nl}(t)|^2\rangle_{\rm en} f_{pp}^{nl}
\mathcal{B}_{nl\rightarrow J/\psi}\over \sum_{nl}
\langle |c_{nl}(t_0)|\rangle^2\rangle_{\rm en}  f_{pp}^{nl} \mathcal{B}_{nl\rightarrow J/\psi}}
\end{align}
where $\mathcal{B}_{nl\rightarrow J/\psi}$ is the branching ratio of charmonium 
eigenstates with the quantum number $(n,l)$ decaying into the ground state $J/\psi$. 
We consider the decay channels of $\chi_c\rightarrow J/\psi$ 
and $\psi(2S)\rightarrow J/\psi$.
$f_{pp}^{nl}$ is the direct production of charmonium eigenstate ($J/\psi$, $\chi_c$, 
$\psi(2S)$) without the feed-down process in pp 
collisions. The ratio of the charmonium direct production is 
extracted to be $f_{pp}^{J/\psi}:f_{pp}^{\chi_c}:f_{pp}^{\psi(2S)}
=0.68:1:0.19$~\cite{ParticleDataGroup:2018ovx}.

\subsection{In-medium heavy quark potential}
\begin{figure}[!hbt]
\centering
\includegraphics[width=0.33\textwidth]{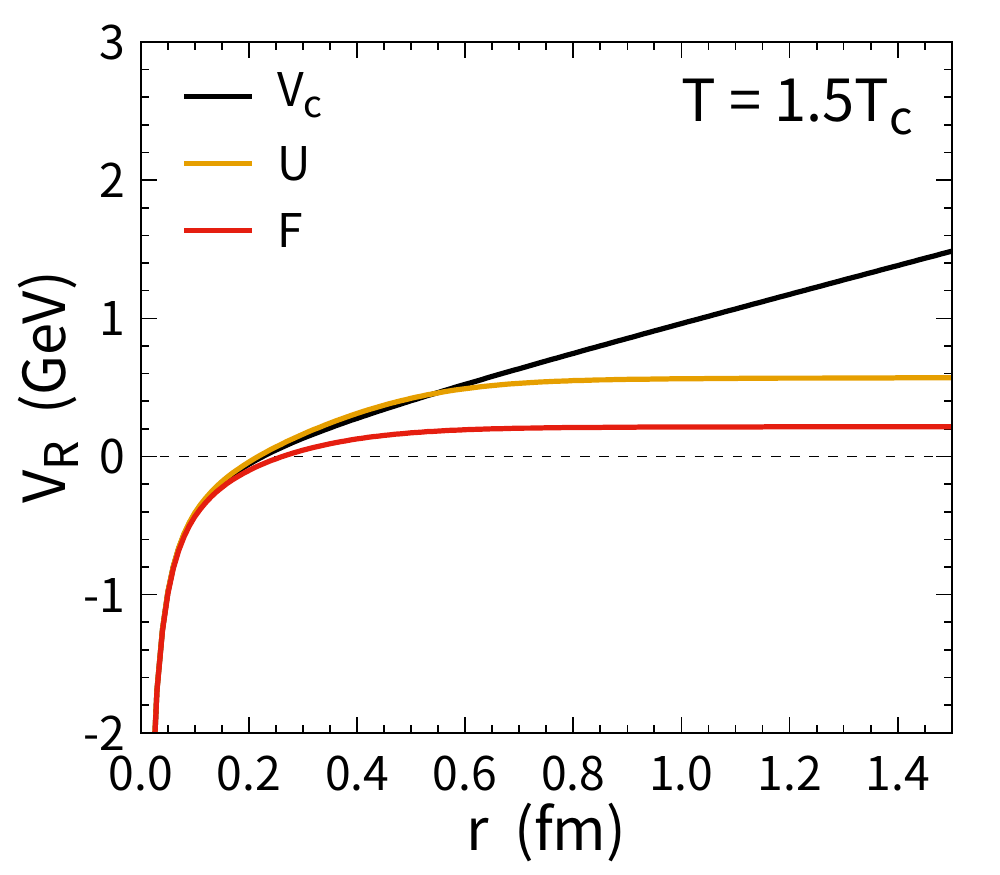}
\caption{(Color online) Different parametrizations of the real part of 
heavy quark potentials as a function of $r$ at 
$T=1.5T_c$. The free energy $F(r,T)$,  
internal energy $U(r,T)$ and the Cornell potential $V_c(r)$ are plotted with different 
color lines. 
}
\hspace{-0.1mm}
\label{fig-diffV}
\end{figure}

In vacuum, heavy quark potential in the quarkonium can be 
approximated as the Cornell potential. At finite temperature, the 
Cornell potential is screened by thermal light partons. 
The real part of in-medium heavy quark potential is between the limits of the free 
energy $F$ and the internal energy $U$ of charmonium. 
The in-medium potential has been studied by Lattice QCD calculations 
and the potential models~\cite{Kaczmarek:2005ui,Digal:2005ht,Shi:2021qri,Lafferty:2019jpr}. 
We parametrize the temperature and coordinate dependence 
of the free energy with the formula, 
\begin{align}
\label{fun-latreal}
F(T,r) =& -{\alpha\over r}[e^{-\mu r}+\mu r]\nonumber \\
& -{\sigma \over 2^{3/4}\Gamma[3/4]}({r\over \mu})^{1/2} K_{1/4}[(\mu r)^2] 
+{\sigma\over 2^{3/2}\mu }{\Gamma[1/4]\over \Gamma[3/4]} 
\end{align}
where 
$\alpha=\pi/12$ and $\sigma=0.2\ \rm{GeV^2}$ are given in  
the Cornell potential $V_c(r)={-\alpha/r}+\sigma r$. 
The $\Gamma$ and $K_{1/4}$ are the Gamma function and the modified Bessel 
function, respectively. 
The screened mass in Eq.~(\ref{fun-latreal}) is taken as~\cite{Digal:2005ht}, 
\begin{align}
{\mu(\bar T)\over \sqrt{\sigma}} = s\bar{T} +a \sigma_t \sqrt{\pi \over 2} 
[\mathrm{erf}({b\over \sqrt{2}\sigma_t}) - \mathrm{erf}({b-\bar{T}\over \sqrt{2}\sigma_t})]
\end{align}
with ${\bar T}\equiv T/T_c$, where $T_c$ is the critical temperature of the 
deconfined phase transition. Other parameters are taken as 
$s=0.587$, $a=2.150$, $b=1.054$, $\sigma_t=0.07379$. 
$\mathrm{erf}(z)$ is the error function. 
The internal energy of heavy quarkonium can be obtained via the relation 
$U(T,r)=F+T(-\partial F/\partial T)$. When the slope of the line  
becomes flat, it indicates that 
there is no attractive force to restraint the wave function at this distance $r$. 
At temperatures around $T_c$,  
there is a sudden shift in the screened mass 
$\mu(\bar T)$~\cite{Digal:2005ht}. The internal energy 
may become slightly larger than the vacuum Cornell potential. 
This behavior can be seen in $U(T,r)$ at $r\sim 0.4$ fm in Fig.~\ref{fig-diffV} 
and become more evident at $T\rightarrow T_c$. 
To avoid this subtlety, we take heavy quark potential 
to be the free energy as the limit of strong color screening, and the vacuum Cornell 
potential as the limit of extremely weak color screening. The realistic potential is 
between these two limits. 
Different 
heavy quark potentials in Fig.~\ref{fig-diffV} will be taken 
into the Schr\"odinger equation to calculate the 
nuclear modification factors of $J/\psi$ and $\psi(2S)$ in the next section.

\begin{figure}[!hbt]
\centering
\includegraphics[width=0.33\textwidth]{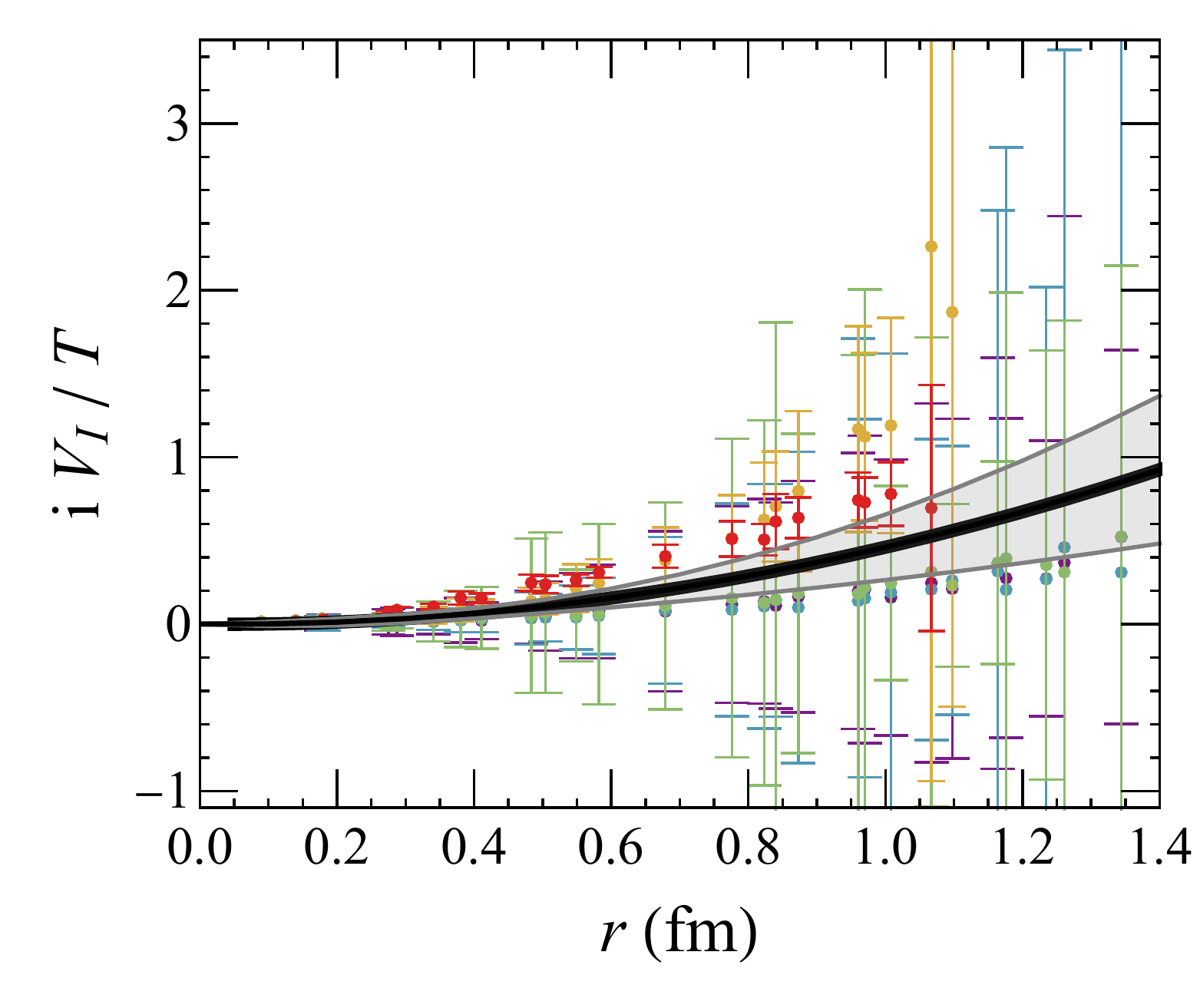}
\caption{(Color online) The imaginary part of the heavy quark potential as a function of distance. The gray band represents the $95\%$ confidence region whereas the black curve corresponds to the 
Maximum Posteriori parameter set.
The data is cited from \cite{Burnier:2016mxc}. Symbols from purple to red correspond to results from low to high temperature. 
}
\hspace{-0.1mm}
\label{lab-fig-imagV}
\end{figure}

In the hot medium, quarkonium bound states can also be dissociated by the inelastic 
scatterings with thermal light partons. This process 
contributes an imaginary part in the 
potential $V(T,r)$. 
We parameterize the temperature and spatial dependence of the imaginary potential by 
\begin{align}
\label{eq-imagV}
&V_I(T,\bar r)= -i\,T(a_1\, {\bar r} + a_2 {\bar r}^2)\,,
\end{align}
where $i$ is the imaginary unit, and
$\bar r\equiv r/{\rm fm}$ is a dimensionless variable. The dimensionless coefficients, $a_1$ and $a_2$, are obtained by invoking the Bayesian inference to fit the lattice QCD calculations~\cite{Burnier:2016mxc}. We focused on the temperature relevant to p-Pb collision $T_c <T< 1.9~T_c$. Results are shown in Fig.~\ref{lab-fig-imagV}, where the gray band represents the $95\%$ confidence interval, and the black curve corresponds to the parameter set $a_1=-0.040$ and $a_2=0.50$ which maximizes the Posterior distribution. In $V_I$, 
the magnitude of the imaginary potential becomes smaller at smaller distance. 
This results in a weaker 
reduction on the $J/\psi$ component than $\psi(2S)$ component 
in the wave function of $c\bar c$ 
dipole. { 
As the imaginary potential in Fig.\ref{lab-fig-imagV} is calculated 
in the gluonic medium, 
we take the same formula in the quark-gluon plasma in heavy-ion 
collisions, which contributes some uncertainty in the suppression of charmonium 
in p-Pb collisions~\cite{Lafferty:2019jpr,Burnier:2014ssa}. 
 The uncertainty of the imaginary potential is partially considered with the theoretical 
band in Fig.\ref{lab-fig-imagV}, which will be reflected in the charmonium $R_{pA}$. 
}

{
In the hot medium produced in p-Pb collisions, heavy quark dipoles 
experience different local temperatures when they move along different trajectories. 
The real and imaginary parts of the potential depending on the 
local temperatures also changes with time. The wave package at each time step 
is obtained from the Schr\"odinger equation, while its normalization is reduced 
by the imaginary part of the Hamiltonian. Therefore, the fractions of charmonium 
eigenstates in the wave package are changed with time 
due to the in-medium potentials. 
}
\vspace*{0pt}
\subsection{Hot medium evolution in p-Pb collisions}

The dynamical evolution of the hot medium produced in 
p-Pb collisions at $\sqrt{s_{NN}}=5.02$ TeV is 
described by the hydrodynamic equations~\cite{Zhao:2020wcd}, 
\begin{align}
\partial_{\mu\nu} T^{\mu\nu}=0
\end{align}
where $T^{\mu\nu}=(e+p)u^\mu u^\nu-g^{\mu\nu}p$ is the energy-momentum tensor. 
$e$ and $p$ is the energy density and the pressure respectively. 
$u^\mu$ is the four velocity of the medium. The equation of state is needed to close 
the hydrodynamic equations. The deconfined phase is treated as an ideal gas of 
gluons and massless $u$ and $d$ quarks plus $s$ quarks with the 
mass $m_s=150$ MeV. 
The confined phase is treated with Hadron Resonance Gas model (HRG)~\cite{Sollfrank:1996hd}. 
Two phases are connected with a first-order 
phase transition. The critical temperature of 
the phase transition is determined to be $T_c=165$ MeV 
by choosing the mean field repulsion 
parameter and the bag constant to be $K=450 \ \rm{MeV\,fm^3}$ 
and $B^{1/4}=236$ MeV~\cite{Zhu:2004nw}. 
{ 
With the multiplicity of light hadrons measured in p-Pb collisions and the 
theoretical simulations from other hydrodynamic models~\cite{ALICE:2014xsp,Zhao:2020wcd},   }
we take 
the maximum initial temperature of the hot medium 
to be $T_0({\bf x}_T=0|b=0)=248$ MeV in forward rapidity 
and $289$ MeV in backward rapidity, respectively. 
{Event-by-event fluctuations 
in the hydrodynamic evolutions are not included yet.}
The profile of the initial energy density is also consistent with 
the 
results from a multiple phase transport (AMPT) model~\cite{Liu:2013via}. 

\begin{figure}[!hbt]
\centering
\includegraphics[width=0.33\textwidth]{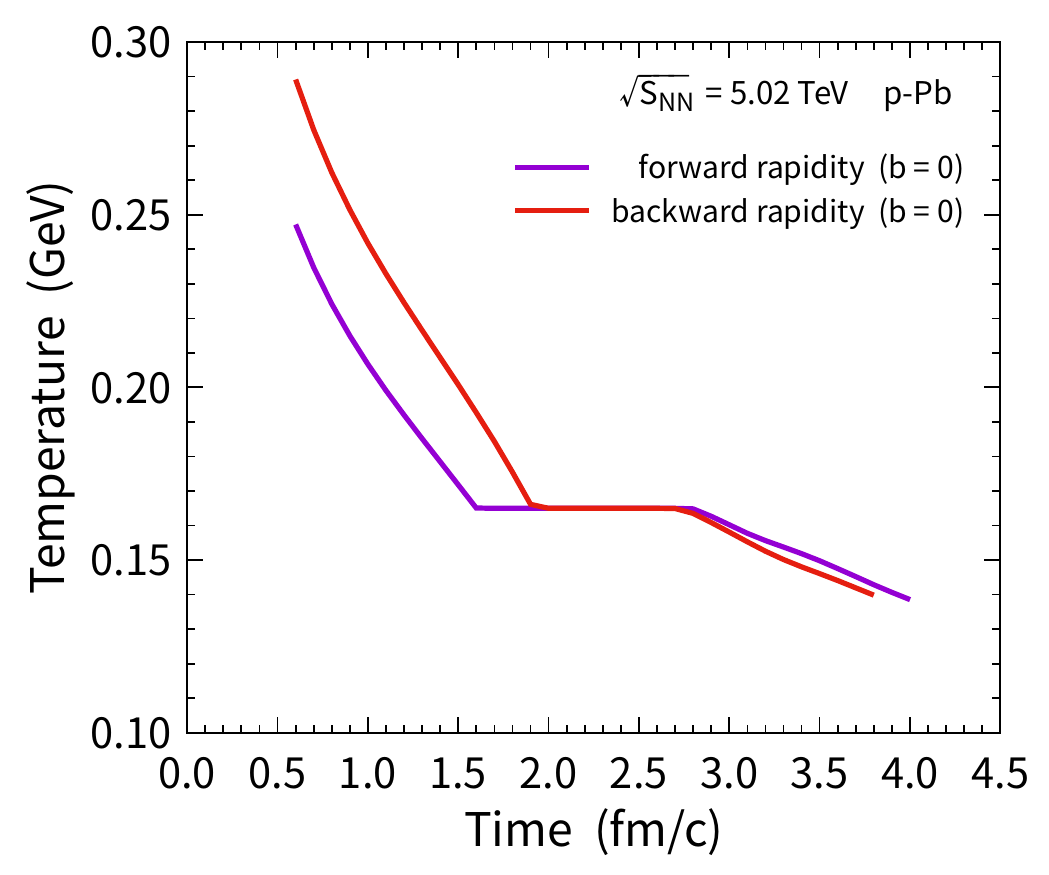}
\caption{(Color online) Time evolution of the temperature in the center 
of the medium (${x}_T=0$) in forward and backward rapidity in 
$\sqrt{s_{NN}}=5.02$ TeV p-Pb collisions. The impact parameter is  
$b=0$ fm, which is defined to be the distance between the centers of the proton and the 
nucleus. 
}
\hspace{-0.1mm}
\label{fig-hydro-plot}
\end{figure}

Hydrodynamic equations start evolution from $\tau_0=0.6$ fm/c 
where the hot medium is assumed to 
reach local equilibrium. At most central collisions with the impact parameter 
b=0, the time evolution of the local temperature at ${ x}_T=0$ in 
forward and backward rapidity is plotted in Fig.~\ref{fig-hydro-plot}. 
Medium evolutions at other impact parameter can be obtained via the scale of 
initial entropy which depends on 
$N_p(b)$ and $N_{coll}(b)$. 

\section{Applications in p-Pb collisions}\label{sec.3}
\begin{figure}[!hbt]
\centering
\includegraphics[width=0.35\textwidth]{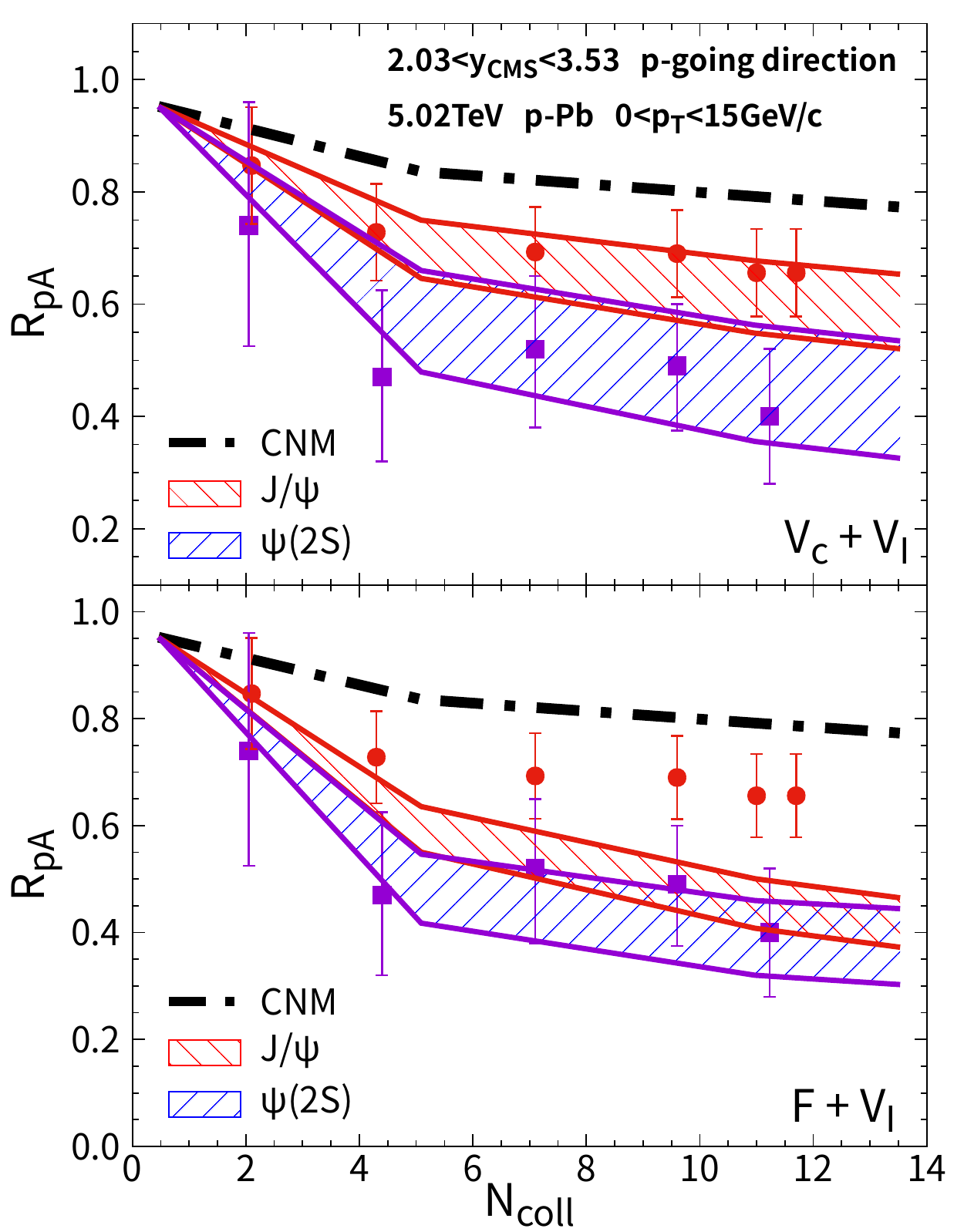}
\caption{(Color online) Nuclear modification factors of $J/\psi$ and $\psi(2S)$ 
as a function of the number of binary collisions $N_{coll}$ in the forward rapidity 
of $\sqrt{s_{NN}}=5.02$ TeV p-Pb collisions. 
Black dashed-dotted line is the calculation 
with only cold nuclear matter effects. The in-medium potential is taken 
to be $V=V_c(r)+V_I(T,r)$ in upper panel and $V=F(T,r)+V_I$ in lower panel. 
Red and blue bands are the results of 
$J/\psi$ and $\psi(2S)$ respectively. 
The experimental data are from the ALICE Collaboration~\cite{ALICE:2015kgk,Leoncino:2016xwa}.
Red circles and blue squares respectively correspond to $J/\psi$ and $\psi(2S)$.
}
\hspace{-0.1mm}
\label{fig-RAA-forwd}
\end{figure}

We apply Schr\"odinger equation to charmonium dynamical evolution in 
$\sqrt{s_{NN}}=5.02$ TeV p-Pb collisions. 
In Fig.~\ref{fig-RAA-forwd}, the {$R_{pA}s$} of
$J/\psi$ and $\psi(2S)$ at forward rapidity (defined as proton-going 
direction) are plotted. The shadowing effect   
modifies the parton densities in the colliding nucleus, which changes the 
gluon density and charmonium production { in nucleus collisions} 
compared with { that} in pp collisions. 
As the shadowing effect exists before the initial production of heavy quark pair in 
parton hard scatterings, it gives the same modification factor  of 
$J/\psi$ and $\psi(2S)$, 
shown as the black 
dotted line in Fig.~\ref{fig-RAA-forwd}. However, the experimental data 
show  different degrees of suppression on the production of 
$J/\psi$ and $\psi(2S)$, which indicates different  strength of  
final state interactions on  different charmonium states. 
The magnitude of the color screening effect on charmonium still deserves 
further investigation. 

\begin{figure}[!hbt]
\centering
\includegraphics[width=0.35\textwidth]{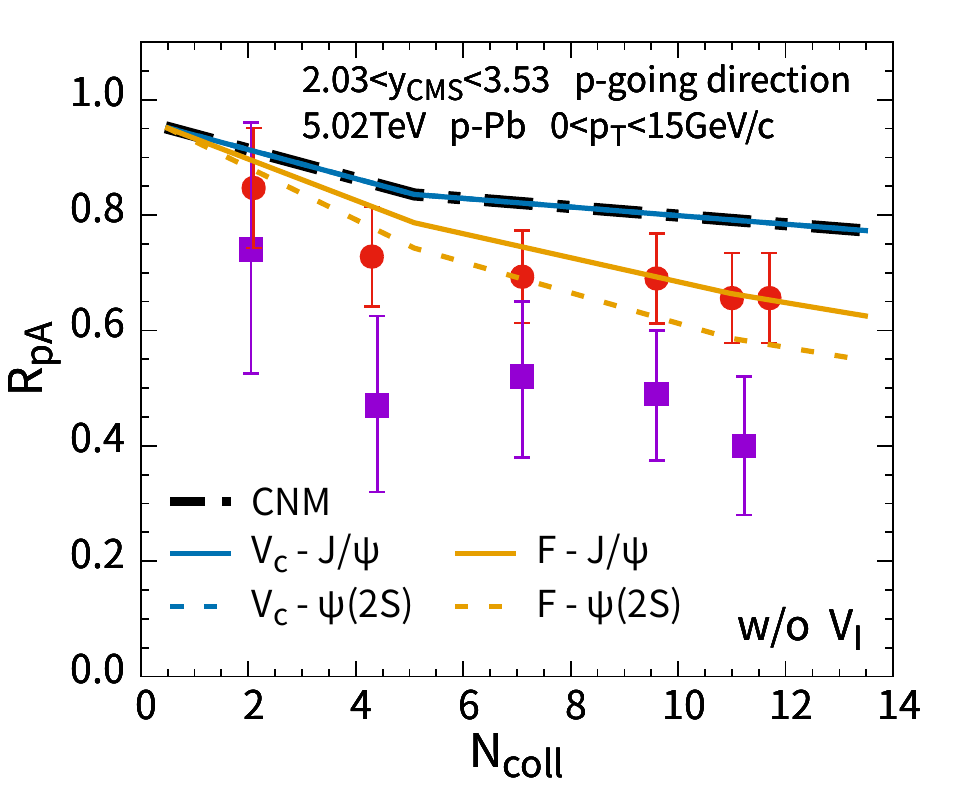}
\caption{(Color online) Nuclear modification factors of $J/\psi$ and $\psi(2S)$ 
as a function of the number of binary collisions $N_{coll}$ in the forward rapidity 
of $\sqrt{s_{NN}}=5.02$ TeV p-Pb collisions. 
 Only real part of the heavy quark potential is included. Black dashed-dotted line is the calculation 
with only cold nuclear matter effects.The strong and weak limits of the potential are taken as the vacuum Cornell potential $V=V_c(r)$ and the free energy $V=F(r,T)$ 
respectively. 
The experimental data are from the ALICE Collaboration~\cite{ALICE:2015kgk,Leoncino:2016xwa}.
Red circles and blue squares respectively correspond to $J/\psi$ and $\psi(2S)$.
}
\hspace{-0.1mm}
\label{fig-RAA-realV}
\end{figure}

In order to study the color screening effect on charmonium observables, we 
  firstly test the scenario without imaginary potential  in Fig.\ref{fig-RAA-realV}.  The calculations with a strong screening scenario with F potential and a weak screening scenario with vacuum potential are presented in the figure. In the strong color screening scenario with F potential, the wave function of 
  $c\bar c$ dipole expands outside due 
  to the weak attractive force between 
  $c$ and $\bar c$. This reduces the 
  overlap of wave-function between $c\bar{c}$ pair and charmonium eigenstate. This suppresses the $R_{pA}$ of $J/\psi$ and $\psi(2S)$. The color screening effect is not strong enough to explain the strong suppression of $\psi(2S)$ $R_{pA}$, which indicates the necessity of  including the imaginary potential in a phenomenological perspective.

In Fig.\ref{fig-RAA-forwd}, both color screened   real potential and imaginary potential are included. In the upper panel of Fig.\ref{fig-RAA-forwd}, only imaginary potential is considered without color screening effect. 
The theoretical band in $R_{pA}$ represents 
the uncertainty in the parametrization of $V_I$. As one can see, the imaginary potential can explain well both $R_{pA}$s of $J/\psi$ and $\psi(2S)$. 
Lower $R_{pA}$ corresponds to 
the upper limit of the $V_I$ parametrization .  As the magnitude of $V_I$ 
increases with the distance, $\psi(2S)$ component in the $c\bar c$ dipole wave function 
is more suppressed. 
In the lower panel of Fig.~\ref{fig-RAA-forwd}, in a strong screening scenario, 
 the real part of heavy quark potential is taken 
as a free energy $V_R=F(T,r)$.

Charmonium wave function is loosely bound 
in the  $c\bar{c}$ wave function. The wave function expands outside, which reduces 
the overlap of wave function between $c\bar c$ wave package and the $J/\psi$ eigenstate.  This results in a transition {  of the final yields} from $J/\psi$ 
to $\psi(2S)$   states and  scattering states. Value of $R_{pA}$ is strongly reduced with 
$V=F+V_I$.     
The feed-down process ($\chi_c,\psi(2S)\rightarrow J/\psi X$) which happens after 
charmonium  escape the hot medium has been included in $R_{pA}$. 
 Comparing the model calculated $R_{pA}s$ with the experimental data,  the vacuum potential is favored and
it seems that the color screening effect is weak for charmonium at the temperatures 
available in p-Pb collisions. 
The imaginary potential is essential to explain 
the difference between $R_{pA}^{J/\psi}$ and $R_{pA}^{\psi(2S)}$  since the real potential in vacuum alone does not change the final projection of the wave-function of $c\bar{c}$ pair to different charmonium species.

\begin{figure}[!hbt]
\centering
\includegraphics[width=0.35\textwidth]{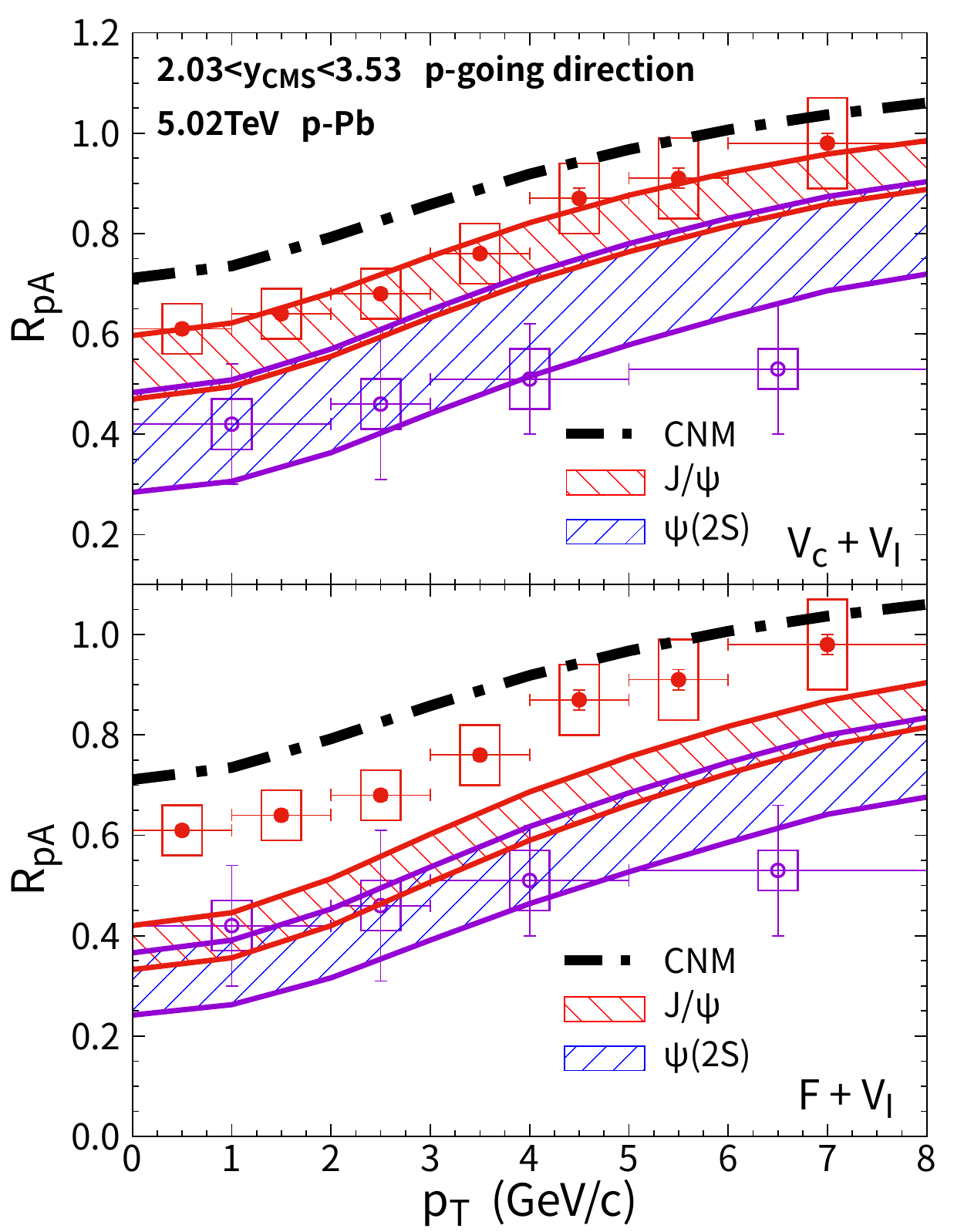}
\caption{(Color online) The $p_T$ dependence of $J/\psi$ and $\psi(2S)$ nuclear 
modification factors in the forward rapidity in minimum-bias $\sqrt{s_{NN}}=5.02$ TeV p-Pb 
collisions. Other conditions are similar to Fig.~\ref{fig-RAA-forwd}. 
The experimental data are from the ALICE Collaboration~\cite{ALICE:2014cgk}. Red circles and blue squares respectively correspond to $J/\psi$ and $\psi(2S)$.
}
\hspace{-0.1mm}
\label{fig-RAApt-forwd}
\end{figure}

The $p_T$ dependence of $J/\psi$ and $\psi(2S)$ $R_{pA}$ is 
also studied in Fig.~\ref{fig-RAApt-forwd}. Black dashed-dotted line is the calculation with 
only cold nuclear matter effects {\color{red}}. 
In the forward rapidity of p-Pb collisions, shadowing effect 
reduce the charmonium production. 
The value of $R_{pA}$ from cold nuclear matter suppression alone increases with transverse momentum due to a weaker shadowing effect at larger transverse energy. 
The dashed-dotted line and the bands increase with $p_T$. 
Besides,   
$c\bar c$ dipoles with large velocities 
move fast  out of the hot medium, where 
$R_{pA}$ becomes larger 
due to the weaker hot medium suppression. 
In the upper panel of Fig.~\ref{fig-RAApt-forwd}, 
theoretical calculations with only imaginary potential 
can explain the $R_{pA}^{J/\psi}$ and $R_{pA}^{\psi(2S)}$ better compared with the 
case of strong color screening effect in the lower panel of Fig.~\ref{fig-RAApt-forwd}. 
The theoretical bands correspond to the uncertainty of $V_I$.

\begin{figure}[hbt]
\centering
\includegraphics[width=0.35\textwidth]{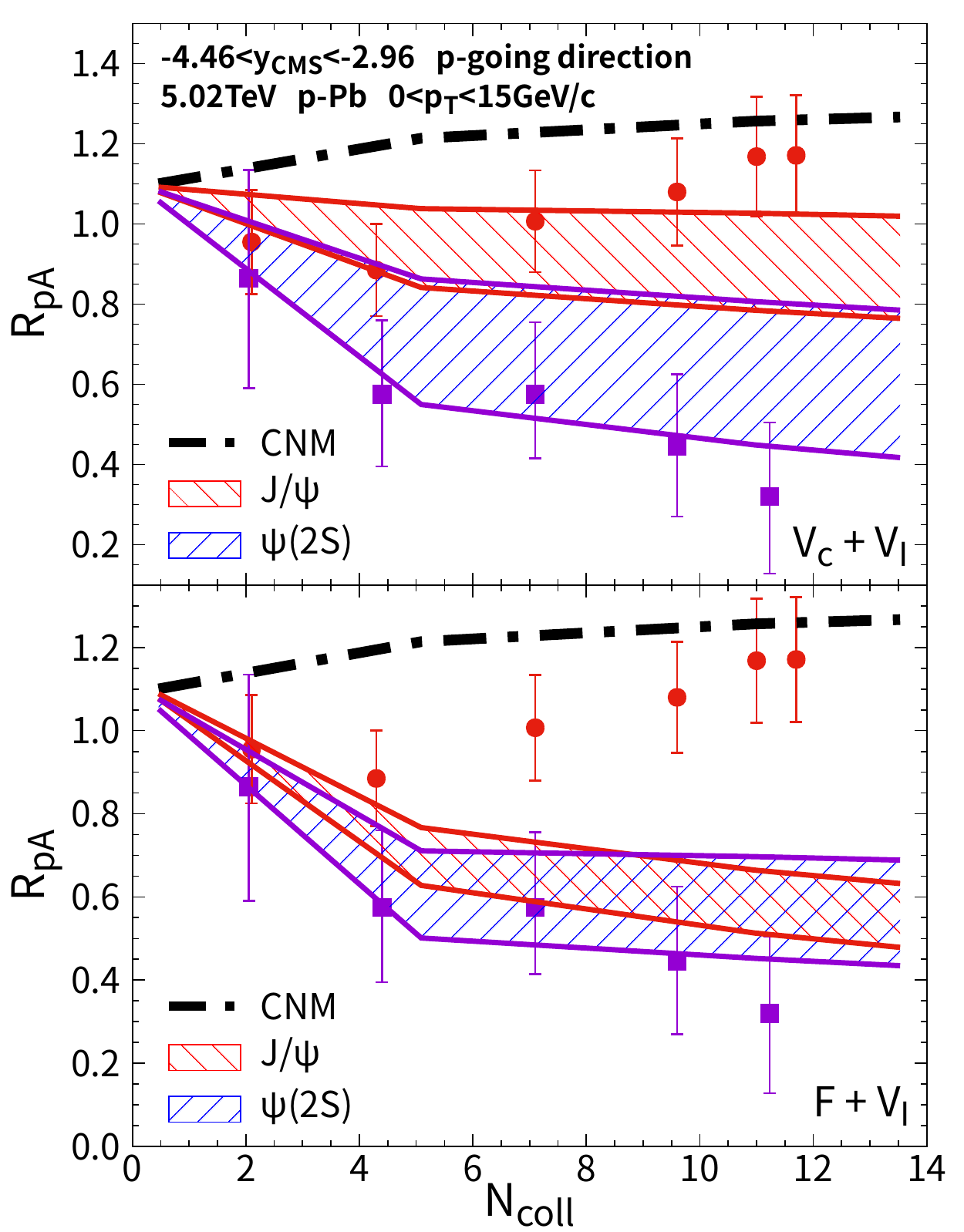}
\caption{(Color online) 
Nuclear modification factors of $J/\psi$ and $\psi(2S)$ 
as a function of the number of binary collisions $N_{coll}$ in the backward rapidity 
of $\sqrt{s_{NN}}=5.02$ TeV p-Pb collisions. 
Red and blue bands are the results of $J/\psi$ and $\psi(2S)$. The bands 
come from the uncertainty of $V_I$. In-medium heavy quark potentials are taken 
as $V=V_c(r)+V_I(T,r)$ in upper panel and $V=F(T,r)+V_I(T,r)$ in lower panel, 
respectively. 
The experimental data are from the ALICE 
Collaboration~\cite{ALICE:2015kgk,Leoncino:2016xwa}.
Red circles and blue squares respectively correspond to $J/\psi$ and $\psi(2S)$.}
\hspace{-0.1mm}
\label{fig-RAAncoll-back}
\end{figure}

In the backward rapidity defined as the Pb-going direction, the anti-shadowing effect 
can increase the $R_{pA}$ of $J/\psi$ and $\psi(2S)$, see the black dashed-dotted line 
in Fig.~\ref{fig-RAAncoll-back}. 
Due to the uncertainty of the 
anti-shadowing effect, 
we consider an upper-limit anti-shadowing effect where the $R_{pA}$ is around 1.27 in most central collisions.  
The $R_{pA}$ with only cold nuclear matter effect is greater than unity. After considering 
the imaginary potential, production of charmonium excited states are 
suppressed, and $R_{pA}^{\psi(2S)}$ is below unity. Since around 40\% of the 
final $J/\psi$ comes from the decay of excited states ($\chi_c$, $\psi(2S)$)  to $J/\psi$, 
the suppression of excited states affect the $R_{pA}^{J/\psi}$ via the feed-down 
process. Shown in the upper panel of Fig.~\ref{fig-RAAncoll-back}, 
 theoretical calculation of $J/\psi$ $R_{pA}$ reproduces the experimental data in peripheral and 
semi-central collisions, while in central collisions $N_{coll}\sim 12$, 
 the theoretical band is at the edge of the experimental data. 
This discrepancy between the theoretical results and the experimental data 
is also reflected in 
semi-classical transport model~\cite{Chen:2016dke} and 
the comover model~\cite{Ferreiro:2014bia}, where $R_{pA}^{J/\psi}\lesssim 1$ 
at $N_{coll}\sim 12$.  
In a strong color screening scenario with F-potential, $J/\psi$ theoretical bands 
strongly underestimate the experimental data. 
This observation is consistent in both backward and forward rapidities. 

\begin{figure}[hbt]
\centering
\includegraphics[width=0.35\textwidth]{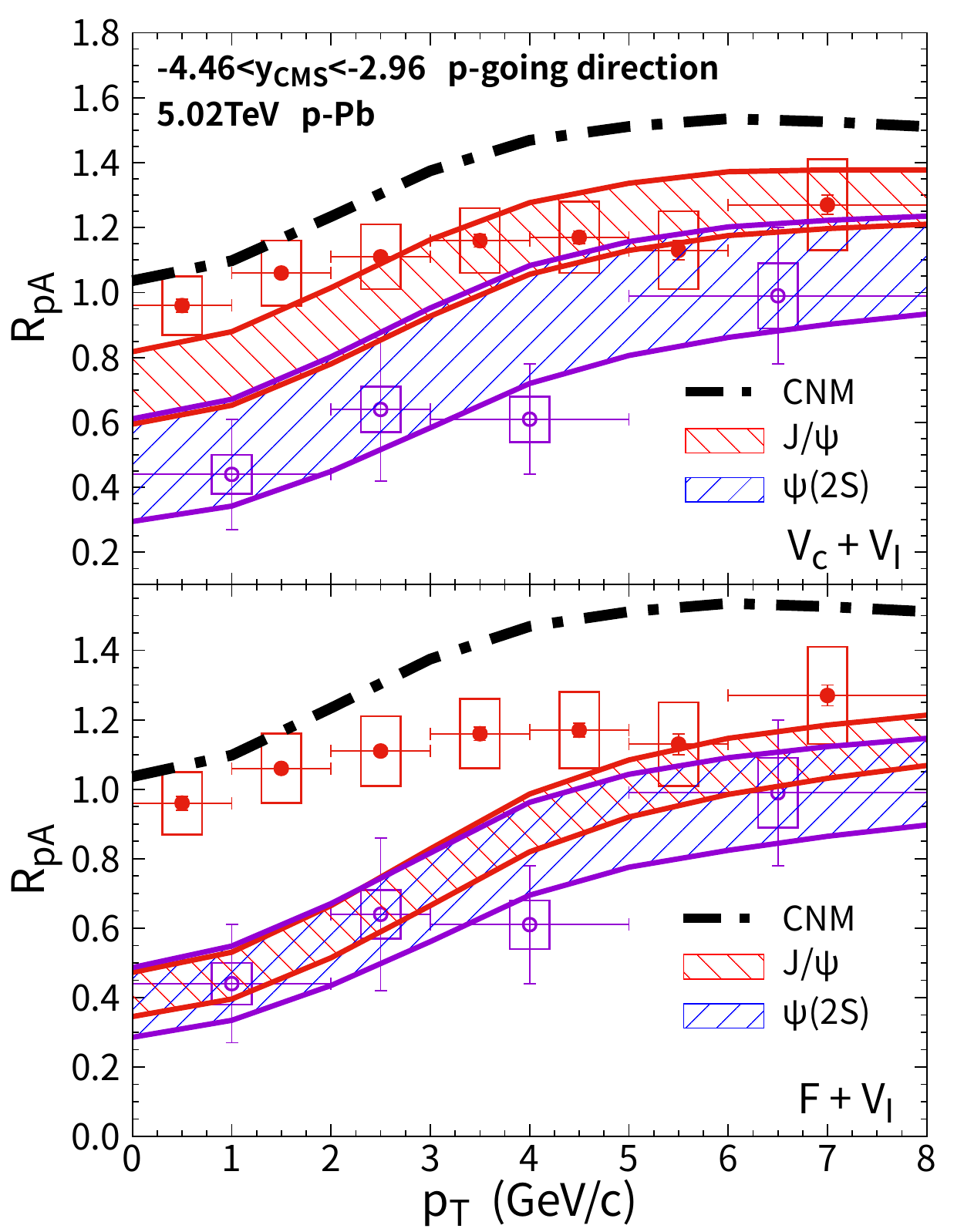}
\caption{(Color online)
The $p_T$ dependence of $J/\psi$ and $\psi(2S)$ nuclear 
modification factors in the backward rapidity in minimum-bias $\sqrt{s_{NN}}=5.02$ TeV p-Pb 
collisions.  
Other conditions are similar to Fig.~\ref{fig-RAAncoll-back}. 
The experimental data are from the 
ALICE Collaboration~\cite{ALICE:2014cgk}. Red circles and blue squares respectively correspond to $J/\psi$ and $\psi(2S)$.}
\hspace{-0.1mm}
\label{fig-RAApt-back}
\end{figure}

The $p_T$ dependence of charmonium $R_{pA}$ is also calculated at the backward rapidity 
 and presented in Fig.~\ref{fig-RAApt-back}. The black dashed-dotted line only includes the 
cold nuclear matter effects. 
Hot medium effects reduce the $R_{pA}$ of $J/\psi$ and $\psi(2S)$ at low $p_T$ region. 
At high $p_T$, anti-shadowing effect make $R_{pA}^{J/\psi}$ become larger than the unity.   
When the real part of the heavy quark potential is taken as the vacuum 
potential, theoretical bands describe the data well in the upper panel of Fig.~\ref{fig-RAApt-back}, while the calculations with F potential in lower panel give small $R_{pA}$ of $J/\psi$ due to the expansion of $c\bar c$ wave package.

\section{conclusion}\label{sec.4}

In this work, we employ a time-dependent Schr\"odinger model to study the 
hot medium effects on charmonium observables in proton-nucleus 
collisions at $\sqrt{s_{NN}}=5.02$~TeV. 
We initialize the $c\bar{c}$ distribution with the cold nuclear matter effects including the (anti-)shadowing effect and the Cronin effect. Both color screening and parton scattering  encodes in the real and the imaginary parts of the potential, which is further incorporated into the Hamiltonian utilized in the quantum evolution. In order to probe the strength of color screening effect, the imaginary part of the potential is constrained by a statistical fit to the lattice QCD data, while two scenarios of the real potential is considered. In the simulation, $c\bar{c}$ dipole initialized with different position and momentum move along different trajectories in the hydrodynamics medium, while their internal evolution is described by the Schr\"odinger equation. 
 The comparison of simulated result with experimental data favors a weak screening scenario, or a strong binding scenario. 
Meanwhile, the imaginary potential is crucial to consistently  describe the
suppression of $J/\psi$ and $\psi(2S)$ states  and the gap between their suppressions due to different width of their wave-functions,  indicating the importance of parton scattering for different charmonium species. 

 The essential phenomenological results from quantum evolution presented in the paper are consistent with those thoroughly studied semi-classical transport approaches.
In a semi-classical approach, the color screening affects the in-medium binding energies of charmonium states, leading to different dissociation widths. While in the potential model discussed in the present work, the color screening broadens the $c\bar{c}$ wave-function, leading to a transfer of bound states to scattering states. The non-Hermitian imaginary part of the potential directly eliminate the tracking of a $c\bar{c}$ pair, corresponding to the dissociation width.
Both effects lead to different suppression strengths but the imaginary part (dissociation) is shown to be crucial for the suppression within both the potential approach and other semi-classical approaches.

There are limitations of this approach as well. Since the size of the $c\bar{c}$ pair is not much smaller than the size of the medium produced in pA collisions, the screening at different positions of the potential could vary. 
Thus a potential model may not be well-defined in this case. However, this approach is one angle of investigating charmonium production in small systems.
A potential model on bottomonium would be favored to study. 
There are statistical extraction of in-medium heavy quark potential with a semi-classical transport approach~\cite{Du:2019tjf} which incorporates the potential in the binding energies and dissociation widths of bottomonium states. Within this potential approach, a direct extraction of the in-medium heavy-quark potential could be performed for bottomonium in AA collisions. We leave that to further publications.

\vspace{1cm}
\appendix {\bf Acknowledgement: }
We appreciate inspiring discussions with Pengfei Zhuang, Ralf Rapp, Shuai Liu and Yunpeng Liu.  
This work is supported by the
National Natural Science Foundation of China (NSFC)
under Grant Nos. 12175165, 11705125. 
S.S. acknowledges support from U.S. Department of Energy, Office of Science, Office of Nuclear Physics, grant No. DE-FG88ER40388. X.D acknowledges support by the Deutsche Forschungsgemeinschaft (DFG, German Research Foundation) through the CRC-TR 211 `Strong-interaction matter under extreme conditions’– project number 315477589 – TRR 211.



\bibliography{ref}

\end{document}